\documentstyle[12pt]{article}

\newcommand{\resection}[1]{\setcounter{equation}{0}\section{#1}}

\newcommand{\EQ}{\begin{equation}}
\newcommand{\EN}{\end{equation}}
\newcommand{\bea}{\begin{eqnarray}}
\newcommand{\eea}{\end{eqnarray}}
\newcommand{\hs}{\hspace{0.1cm}}

\newcommand{\be}{\beta}

\begin{document}
\setcounter{page}{0}
\topmargin 0pt
\oddsidemargin 5mm
\renewcommand{\thefootnote}{\fnsymbol{footnote}}
\newpage
\setcounter{page}{0}
\begin{titlepage}
\begin{flushright}
ISAS/EP/92/208 \\
USP-IFQSC/TH/92-51
\end{flushright}
\vspace{0.5cm}
\begin{center}
{\large {\bf Form Factors of the Elementary Field in the Bullough-Dodd
Model}} \\
\vspace{1.8cm}
{\large A. Fring}\\
\vspace{0.5cm}
{\em Universidade de S\~ao Paulo, \\
Caixa Postal 369, CEP 13560 - S\~ao Carlos-SP,
Brasil }\\
\vspace{1cm}
{\large G. Mussardo, P. Simonetti}\\
\vspace{0.5cm}
{\em International School for Advanced Studies,\\
and \\
Istituto Nazionale di Fisica Nucleare\\
34014 Trieste, Italy}\\

\end{center}
\vspace{1.2cm}

\renewcommand{\thefootnote}{\arabic{footnote}}
\setcounter{footnote}{0}

\begin{abstract}
We derive the recursive equations for the form factors of the local
hermitian operators in the Bullough-Dodd model. At the self-dual point of
the theory, the form factors of the fundamental field of the Bullough-Dodd
model are equal to those of the fundamental field of the Sinh-Gordon model at
a specific value of the coupling constant.
\end{abstract}
%\vspace{.3cm}
\end{titlepage}

\newpage

\resection{The Bullough-Dodd Model}

The Bullough-Dodd (BD) model \cite{DB,ZS} is an integrable quantum field
theory defined by the Lagrangian
\EQ
{\cal L}\,=\,\frac{1}{2}(\partial_{\mu}\varphi)^2-
\frac{m^2_0}{6g^2}\left(2e^{g\varphi}+e^{-2g\varphi}\right)\,\, ,
\label{BuD}
\EN
where $m_0$ is a bare mass parameter and $g$ the coupling constant. This model
belongs to the class of the Affine Toda Field Theories\footnote{Classically,
this model can be obtained by a $Z_2$ folding of the simply-laced $A_2$ Affine
Toda theory.} \cite{MOP} corresponding to the non-simply laced root system
$BC_1$ with Coxeter number $h=3$. The expansion of the exponential terms in
(\ref{BuD}) gives rise to the $n$-point interaction vertices
\EQ
V(\varphi)\,=\,\frac{m_0^2}{6g^2}\left(2e^{g\varphi}+
e^{-2g\varphi}\right)\,=\,\frac{m_0^2}{2\,g^2}+
\sum_{k=2}^{\infty}\frac{g_k}{k!}\varphi^k\,\, ,
\EN
where
\EQ
g_k\,=\,\frac{m_0^2}{3}g^{k-2}\left[1+(-1)^k 2^{k-1}\right]\,\, .
\EN
The infinite renormalization of the BD model reduces to a subtraction of the
tadpole diagrams \cite{MC,Musrep}, i.e. the bare mass $m_0$
renormalizes as
\EQ
m^2_0\,\rightarrow m_0^2\left(\frac{\Lambda}{\mu}\right)^{g^2/4\pi}
\EN
where $\Lambda$ is a ultraviolet cut-off and $\mu$ a subtraction point. The
coupling constant $g$ does not renormalize.

It is interesting to note that although the BD model involves a non-simply
laced algebra, it is not plagued by the difficulties which arise in the
analysis of the non-simply laced Toda Field Theories \cite{MC,Musrep}. The
reason essentially lies in the fact that only one field is involved in the
interaction. Therefore, most of the formulae worked out for the simply-laced
Toda Field Theories also apply to the BD model. For instance, the (finite)
wave function and mass renormalization at order one-loop read
\begin{eqnarray}
Z\,&=&\,1-\frac{g^2}{12}\left(\frac{1}{\pi}-\frac{1}{3{\sqrt 3}}\right)
\,\, ,\\
\delta m^2\,&=&\,m^2\frac{g^2}{12{\sqrt 3}}\nonumber\,\, ,
\end{eqnarray}
which coincide with the corresponding expressions obtained for the
simply-laced Toda Field Theories \cite{MC,Musrep}, once we substitute $h=3$.
However, as we will discuss later, the non-simply laced nature of the BD model
manifests itself in a subtle way.

The spectrum of the model consists of a particle state $A$ that appears as
bound state of itself in the scattering process (fig.\,1)
\EQ
A\,\times\, A\,\rightarrow A\,\rightarrow A\,\times\, A \,\,\, .
\label{boot}
\EN
The bootstrap dynamics of the model is supported by an infinite set of local
conserved charges ${\cal Q}_s$ where $s$ is an odd integer but multiple of
$3$
\EQ
s\,=\,1,5,7,11,13,\ldots   \label{eq: spins}
\EN
The absence of spin $s$ multiple of $3$ is consistent with the bootstrap
process (\ref{boot}) \cite{Zam}. The integrability of the BD model implies
the elasticity of the scattering processes \cite{ZZ}. The $n$-particle
$S$-matrix then factorizes into $n(n-1)/2$ two-particle scattering amplitudes,
whose expression is given by \cite{AFZ}
\EQ
S(\beta,B)\,=\,f_{\frac{2}{3}}(\beta)\,f_{\frac{B}{3}-\frac{2}{3}}(\beta)\,
f_{-\frac{B}{3}}(\beta)\,\, ,
\label{BDD}
\EN
where
\EQ
f_x(\beta)\,=\,\frac{\tanh\frac{1}{2}(\beta+i\pi x)}
{\tanh\frac{1}{2}(\beta-i\pi x)}\,\, ,
\EN
and $B(g)$ is the following function of the coupling constant $g$
\EQ
B(g)\,=\,\frac{g^2}{2\pi}\frac{1}{1+\frac{g^2}{4\pi}}\,\,\, .
\EN
The $S$-matrix (\ref{BDD}) is invariant under
\EQ
B(g)\,\rightarrow\, 2-B(g) \,\,\, ,
\EN
i.e. under the weak-strong coupling constant duality
\EQ
g\, \rightarrow \,\frac{4\pi}{g} \,\,\, .
\EN
The minimal part of the $S$-matrix (\ref{BDD}), i.e. $f_{\frac{2}{3}}(\beta)$
coincides with the $S$-matrix of the Yang-Lee model \cite{Cardymus}. For all
values of $g$ except for $0$, $\infty$ and the self-dual point
$g=\sqrt{ 4\pi}$, the $S$-matrix possesses a simple pole at
$\beta=2\pi i/3$, that corresponds to the bound state in the $s$-channel of
the particle $A$ itself. The residue is related to the three-particle vertex
on mass-shell (fig.\,1)
\EQ
\Gamma^2(B)\,=\,2{\sqrt 3} \frac{\tan\left(\frac{\pi B}{6}\right)}
{\tan\left(\frac{\pi B}{6}-\frac{2\pi}{3}\right)}
\frac{\tan\left(\frac{\pi}{3}-\frac{\pi B}{6}\right)}
{\tan\left(\frac{\pi B}{6}+\frac{\pi}{3}\right)}\,\, . \label{eq: threepv}
\EN
Notice that $\Gamma(B)$ vanishes for $B=0$ and $B=2$ (which correspond
to the free theory limits) but it also vanishes at the self-dual point $B=1$
where the $S$-matrix reduces to
\EQ
S(\beta,1)\,=\,f_{-\frac{2}{3}}(\beta) \,\, .
\EN
The vanishing of $\Gamma$ at the self-dual point is the reminiscence
of the non-simply laced nature of this theory. Indeed, the vanishing of
the on-shell three-point vertex at a finite value of the coupling constant
never occurs in any simply-laced Toda Field Theory but may occur for
theories obtained as folding of simply-laced Toda Field Theories \cite{KM}.
We will show later on that at the self-dual point the BD dynamically realizes
an effective $Z_2$ symmetry and we will identify the BD model at the
self-dual point with a specific point of the Sinh-Gordon model.

\resection{Form Factors}

Correlation functions of the BD model may be computed exploiting
the form factor approach [12-19]. The form factors (FF) are matrix elements of
local operators between the vacuum and $n$-particle in-state
\EQ
F_n^{\cal O}(\beta_1,\beta_2,\ldots,\beta_n)\,=\,
<0|{\cal O}(0)|\beta_1,\beta_2,\ldots,\beta_n>_{\rm in}\,\, .
\EN
For local scalar operators ${\cal O}(x)$, relativistic invariance implies
that $F_n$ are functions of the difference of the rapidities. Except for the
poles corresponding to the one-particle bound states in all sub-channels, we
expect the form factors $F_n$ to be analytic inside the strip
$0 < {\rm Im } \be_{ij} < 2\pi$. In this paper we consider the form
factors of the fundamental field $\varphi(x)$ of the BD model. Form
factors of general operators of the BD model are analyzed in a forthcoming
paper \cite{FMS3}.

The form factors of the hermitian local scalar operator ${\cal O}(x)$ satisfy
a set of functional equations, known as Watson's equations \cite{Watson},
which for integrable systems assume a particularly simple form
\bea
F_n(\be_1, \dots ,\be_i, \be_{i+1}, \dots, \be_n) &=& F_n(\be_1,\dots,\be_{i+1}
,\be_i ,\dots, \be_n) S(\beta_i-\beta_{i+1}) \,\, ,
\label{permu1}\\
F_n(\be_1+2 \pi i, \dots, \be_{n-1}, \be_n ) &=& F_n(\be_2 ,\ldots,\be_n,
\be_1) = \prod_{i=2}^{n} S(\beta_i-\beta_1) F_n(\be_1, \dots, \be_n)
\,\, .
\nonumber
\eea
In the case $n=2$, eqs.\,(\ref{permu1}) reduce to
\EQ
\begin{array}{ccl}
F_2(\beta)&=&F_2(-\beta)S_2(\beta) \,\, ,\\
F_2(i\pi-\beta)&=&F_2(i\pi+\beta) \,\,\, .
\end{array}
\label{F2}
\EN
The general solution of Watson's equations can always be brought into the form
\cite{Karowski}
\EQ
F_n(\beta_1,\dots,\beta_n) =K_n(\beta_1,\dots,\beta_n) \prod_{i<j}F_{\rm min}
(\beta_{ij})  \,\, ,
\label{parametrization}
\EN
where $F_{\rm min}(\beta)$ has the properties that it satisfies (\ref{F2}), is
analytic in $0\leq$ Im $\beta\leq \pi$, has no zeros and poles in
$0<$ Im $\beta<\pi$, and converges to a constant value for large values of
$\beta$. These requirements uniquely determine this function, up to a
normalization. In the case of the BD model, $F_{\rm min}(\beta)$ is given by
\begin{eqnarray}
&& F_{\rm min}(\beta,B)\,=\,
\prod_{k=0}^{\infty}
\left|
\frac{\Gamma\left(k+\frac{3}{2}+\frac{i\hat\beta}{2\pi}\right)
\Gamma\left(k+\frac{7}{6}+\frac{i\hat\beta}{2\pi}\right)
\Gamma\left(k+\frac{4}{3}+\frac{i\hat\beta}{2\pi}\right)}
{\Gamma\left(k+\frac{1}{2}+\frac{i\hat\beta}{2\pi}\right)
\Gamma\left(k+\frac{5}{6}+\frac{i\hat\beta}{2\pi}\right)
\Gamma\left(k+\frac{2}{3}+\frac{i\hat\beta}{2\pi}\right)}
\right.
\\
&&\,\,\,\times\,\,\left.
\frac{\Gamma\left(k+\frac{5}{6}-\frac{B}{6}+\frac{i\hat\beta}{2\pi}\right)
\Gamma\left(k+\frac{1}{2}+\frac{B}{6}+\frac{i\hat\beta}{2\pi}\right)
\Gamma\left(k+1-\frac{B}{6}+\frac{i\hat\beta}{2\pi}\right)
\Gamma\left(k+\frac{2}{3}+\frac{B}{6}+\frac{i\hat\beta}{2\pi}\right)}
{\Gamma\left(k+\frac{7}{6}+\frac{B}{6}+\frac{i\hat\beta}{2\pi}\right)
\Gamma\left(k+\frac{3}{2}-\frac{B}{6}+\frac{i\hat\beta}{2\pi}\right)
\Gamma\left(k+1+\frac{B}{6}+\frac{i\hat\beta}{2\pi}\right)
\Gamma\left(k+\frac{4}{3}-\frac{B}{6}+\frac{i\hat\beta}{2\pi}\right)}
\nonumber \right|^2
\end{eqnarray}
with $\hat\beta=i\pi-\beta$. $F_{\rm min}(\beta,B)$ has a simple zero at the
threshold $\beta=0$ (since $S(0,B)=-1$) and its asymptotic behaviour is given
by
\EQ
\lim_{\beta \rightarrow \infty} F_{\rm min}(\beta,B) = 1\,\,\, .
\EN
It satisfies the functional equations
\begin{eqnarray}
F_{\rm min}(i\pi+\beta) F_{\rm min}(\beta)\,&=&\,
\frac{\sinh\beta\left(\sinh\beta+\sinh\frac{i\pi}{3}\right)}
{\left(\sinh\beta+\sinh\frac{i\pi B}{3}\right)
\left(\sinh\beta+\sinh\frac{i\pi(1+B)}{3}\right)}
\nonumber\\
F_{\rm min}(\beta+\frac{i\pi}{3}) F_{\rm min}(\beta-\frac{i\pi}{3})\,&=&\,
\frac{\cosh\beta+\cosh\frac{2i\pi}{3}}
{\cosh\beta+\cosh\frac{i\pi(2+B)}{3}}
\,\,F_{\rm min}(\beta) \label{shift}
\end{eqnarray}
which are quite useful in the derivation of the recursive equations
for the form factors.

The remaining factors $K_n$ in (\ref{parametrization}) then satisfy Watson's
equations with $S_2=1$, which implies that they are completely symmetric,
$2 \pi i$-periodic functions of the $\beta_{i}$. They must contain all the
physical poles expected in the form factor under consideration and must
satisfy a correct asymptotic behaviour for large value of $\beta_i$.
{}From the LSZ-reduction \cite{FMS}, the form factors of the elementary field
$\varphi(0)$ behave asymptotically as
\EQ
F_n(\beta_1,\ldots,\beta_n)\,\rightarrow 0
\,\,\,\,\,
{\rm as} \,\, \beta_i\rightarrow +\infty\,\,\,\,\,\beta_{j\neq i}\,\,\,\,\,
{\rm fixed}
\,\,\, (n>1)\,\,\, ,
\EN
with the normalization given by
\EQ
<0|\varphi(0)|\beta>\,=\,\frac{1}{\sqrt 2} \,\,\, ,
\EN
i.e. $\varphi(0)$ creates a one-particle state from the vacuum.
Taking into account the bound state pole in the two-particle channel at
$\beta_{ij}=2\pi i/3$ and the one-particle pole in the three-particle channel
at $\beta_{ij}=i\pi$, the general form factors can be parameterized as
\EQ
F_n(\beta_1,\ldots,\beta_n)\,=
%\, H_n
\, Q_n(x_1,\ldots,x_n)\,
\prod_{i<j} \frac{F_{\rm min}(\beta_{ij})}{(x_i+x_j)(\omega x_i+x_j)
(\omega^{-1}x_i+x_j)} \,\, , \label{para}
\EN
where we have introduced the variables
\EQ
x_i\,=\,e^{\beta_i} \,\, , \,\, \omega\,=\,e^{i\pi/3}\,\, .
\EN
%and the normalization constant $H_n$.
The functions $Q_n(x_1,\dots,x_n)$ are symmetric polynomials\footnote{The
polynomial nature of the functions $Q_n$ is dictated by the locality of the
theory \cite{nankai}.} in the variables $x_i$. They can be expressed in terms
of {\em elementary symmetric polynomial} $\sigma^{(n)}_k(x_1,\dots,x_n)$ which
are generated by \cite{Macdon}
\EQ
\prod_{i=1}^n(x+x_i)\,=\,
\sum_{k=0}^n x^{n-k} \,\sigma_k^{(n)}(x_1,x_2,\ldots,x_n) \,\,\, .
\label{generating}
\EN
As proved in \cite{FMS}, a convenient parametrization of the polynomials
$Q_n$ entering the form factors of the elementary field is given by
\EQ
Q_n(x_1,x_2,\ldots,x_n)\,=\,
\sigma_n^{(n)}\,P_n(x_1,x_2,\ldots,x_n) \,\,\, ,
\label{param}
\EN
where $P_n$ are symmetric polynomials of total degree $n(3n-5)/2$ and of
degree $3n-5$ in each variable $x_i$. The explicit determination of the
symmetric polynomials $P_n$ is achieved by means of the recursive equations
satisfied by the form factors.

\subsection{Pole Structure and Residue Equations for the Form Factors}

The pole structure of the form factors induces a set of recursive equations
for the $F_n$. The first kind of poles arises from kinematical poles located at
$\beta_{ij}=i\pi$. The corresponding residues are computed by the LSZ
reduction \cite{Smirnov1,Smirnov2} and give rise to a recursive equation
between the $n$-particle and the $(n+2)$-particle form factors
\EQ
-i\lim_{\tilde\beta \rightarrow \beta}
(\tilde\beta - \beta)
F_{n+2}(\tilde\beta+i\pi,\beta,\beta_1,\beta_2,\ldots,\beta_n)=
\left(1-\prod_{i=1}^n S(\beta-\beta_i)\right)\,
F_n(\beta_1,\ldots,\beta_n)  . \label{recursive}
\EN
For the BD model, using the parameterization (\ref{para}) this
equation becomes
\EQ
(-1)^{n} Q_{n+2}(-x,x,x_1,x_2,\ldots,x_n)\,=
\,\frac{1}{F_{\rm min}(i\pi)}\,x^3 \,U(x,x_1,x_2,\ldots,x_n)
Q_n(x_1,x_2,\ldots,x_n)
\EN
where
\bea
U(x,x_1,\ldots,x_n) &=& 2\sum_{k_1,\ldots,k_6=0}^n
(-1)^{k_2+k_3+k_5}\, x^{6n-(k_1+\cdots\, +k_6)}
 \,\sigma_{k_1}^{(n)}\sigma_{k_2}^{(n)}\ldots\sigma_{k_6}^{(n)} \, \\
&&\,\,\, \times\,
\sin\left[\frac{\pi}{3}\left[2(k_2+k_4-k_1-k_3)+B(k_3+k_6-k_4-k_5)\right]
\right] .\nonumber
\eea
This equation establishes a recursive structure between the $(n+2)$- and
$n$-particle form factors.

The second type of poles in the $F_n$ arises from the bound state singularity.
The corresponding residue for the $F_n$ is given by
\cite{Smirnov1,Smirnov2}
\EQ
-i\lim_{\epsilon\rightarrow 0} \epsilon\,
F_{n+1}(\beta+i \frac{\pi}{3}-\epsilon,
\beta-i \frac{\pi}{3}+\epsilon,\beta_1,\ldots,\beta_{n-1})
\,=\,\Gamma(g)\,F_{n}(\beta,\beta_1,\ldots,\beta_{n-1})
\,\,\, ,
\label{respole}
\EN
For the BD model, eq.\,(\ref{respole}) becomes
\EQ
Q_{n+2}(\omega x,\omega^{-1}x,x_1,\ldots,x_n)\,=\,-\,
\frac{\sqrt{3}}{F_{\rm min}(i \frac{2\pi}{3})} \,
\Gamma(g) \,x^3 D(x,x_1,\ldots,x_n) \,Q_{n+1}(x,x_1,\ldots,x_n)
\EN
where
\bea
D(x,x_1,\ldots,x_n)\,&=&\,
\prod_{i=1}^n (x+x_i)(x \omega^{2+B} + x_i)
(x \omega^{-B-2} + x_i ) \\
 &=&\,
\sum_{k_1,k_2,k_6=0}^n
x^{3n-(k_1+k_2+k_6)} \, \omega^{(2+B)(k_2-k_3)} \,
 \,\sigma_{k_1}^{(n)}\sigma_{k_2}^{(n)}\sigma_{k_3}^{(n)} \,\,\,.
\nonumber
\eea
This equation establishes a recursive structure between the $(n+2)$- and
$(n+1)$-particle form factors.

In this paper we mainly focalize our attention on the solution of the
recursive equations at the self-dual point.

\resection{BD Model at the Self-Dual Point}

The self-dual point of the BD model ($B=1$, i.e. $g=\sqrt{4\pi}$) is a rather
special point in the space of coupling constant. Indeed, at this point two
zeros present in the $S$-matrix move simultaneously to the location of the
pole and cancel it. Thus the $S$-matrix reduces to
\EQ
S(\beta,B=1)\,=\,f_{-\frac{2}{3}}(\beta) \,\,\, .
\EN
This $S$-matrix is equal to the $S$-matrix of the Sinh-Gordon model,
defined by the Lagrangian
\EQ
{\cal L}\,=\,\frac{1}{2}(\partial_{\mu}\varphi)^2\,-\,
\frac{m_0^2}{\lambda^2} \cosh \lambda\,\varphi(x)\,\, ,
\label{Sinh-Gordon}
\EN
at $\lambda=2\sqrt \pi$ \cite{AFZ}. This equality has a far reaching
consequences. In fact, as we are going to show, not only the scattering
amplitudes of the two theories coincide but also the matrix elements of
the fundamental field of the two theories are the same. This means, in
particular, that all form factors of the elementary field $\varphi(x)$ with
an even number of external legs vanish, whereas the odd ones coincide with
those of the Sinh-Gordon theory at $\lambda=2\sqrt \pi$. This
identity between the two theories is a remarkable equivalence because, looking
at their Lagrangian, the Sinh-Gordon model presents a $Z_2$ invariance while
the BD model apparently not. Hence, in the BD model the $Z_2$ symmetry is
dynamically implemented to the self-dual point.

\subsection{Form Factors}

Next to the one-particle matrix element (that fixes the normalization),
the first non-trivial form factor of the elementary field $\varphi(0)$
is given by the matrix elements on two-particle state. Its expression is
\EQ
F_2(\beta_1,\beta_2)\,=\,-\frac{1}{2}\,\sqrt{\frac{3}{2}}\,
\frac{\Gamma(g)}{\cosh\beta_{12}+\frac{1}{2}}\,
\frac{F_{\rm min}(\beta_{12})}{F_{\rm min}\left(\frac{2\pi i}{3}\right)}
\,\,\, .
\label{twofield1}
\EN
Going to the bound state pole, it correctly reduces to the one-particle form
factor with the residue equal to the three-particle vertex on mass-shell
\EQ
-i\,\lim_{\beta\rightarrow \frac{2\pi i}{3}}\,
\left(\beta-\frac{2\pi i}{3}\right)\,F_2(\beta)\,=\,
\frac{\Gamma(g)}{\sqrt 2} \,\,\, .
\EN
It goes asymptotically to zero, due to the propagator left by the LSZ
reduction. In terms of our parameterization (\ref{para}), we have
\EQ
Q_2(x_1,x_2) \,=\,-\,\sqrt{\frac{3}{2}}\,
\frac{1}{F_{\rm min}\left(\frac{2\pi i}{3}\right)}\,
\Gamma(g) \,\sigma_1(x_1,x_2) \sigma_2(x_1,x_2) \,\,\,.
\EN
The important feature of the two-particle form factor of the renormalized
field $\varphi(0)$ is its proportionality to the three-particle vertex
$\Gamma(g)$. Therefore, at the self-dual point this
form-factor is zero. The vanishing of $F_2$ implies that
of all form factors with even number of external legs are zero
\EQ
F_{2n} \,=\, 0 \,\,\, .
\EN
The proof is given by induction. An important quantity entering the proof
is given by
\EQ
\prod_{1 \leq i<j \leq n} (x_i \omega + x_j) (x_i \omega^{-1} + x_j) =
\det {\cal A} = A^{(n)}  \,\,\, ,
\label{eq: polepr}
\EN
where ${\cal A}$ is a $(2n-2) \times (2n-2)$-matrix with
entries\footnote{$[a]$ is the integer part of the real number $a$.}
\EQ
{\cal A}_{ij}\,=\,\sigma^{(n)}_{3\left[\frac{j}{2}\right]-i+1+(-1)^{j+1}}
\EN
i.e. (suppressing the superscript $(n)$)
\EQ
{\cal A} = \left( \begin{array}{lllll}
\sigma_1 & \sigma_2 & \sigma_4 & \sigma_5 & \cdots \\
       1 & \sigma_1 & \sigma_3 & \sigma_4 & \cdots \\
    0    &  1       & \sigma_2 & \sigma_3 & \cdots \\
    0    &  0       & \sigma_1 & \sigma_2 & \cdots \\
 \vdots  &  \vdots  &  \vdots  &  \vdots  & \ddots \\
\end{array} \right)
\,\,\, .
\EN
Let us consider initially the next form factor with even number of legs, i.e.
$F_4$. With our parametrization, it has to satisfy the following
two recursive equations
\EQ
Q_{4}(\omega x,\omega^{-1}x,x_1,\ldots,x_n)\,=\, 0\,\,\, ,
\label{fir}
\EN
(since $\Gamma$ is zero) and
\EQ
Q_{4}(x,-x,x_1,\ldots,x_n)\,=\, 0\,\,\, ,
\label{sec}
\EN
(since $F_2$ is zero). The general solution of eq.\,(\ref{fir}) is given
in terms of $A^{(4)}$ times a symmetric polynomial $\tilde {\cal S}_4$. This
polynomial can be further factorized as
\EQ
\tilde {\cal S}_4\,=\,\sigma_4 \,{\cal S}_4(x_1,x_2,x_3,x_4) \,\,\, .
\EN
The total degree of ${\cal S}_4$ is $2$ and degree $1$ in each variable. The
only possibility is
\EQ
{\cal S}_4(x_1,\ldots,x_4)\,=\,c\,\sigma_2^{(4)} \,\,\, .
\EN
where $c$ is a constant. But such ${\cal S}_4$ does not satisfy the second
recursive equation (\ref{sec}) unless $c=0$. Hence, in addition to
$F_2$, $F_4$ is also zero at the self-dual point. Using the same kind of
argument we can prove that $F_6$ is also zero at the self-dual point and,
in general, that all $F_{2n}$ vanish.

Concerning the form factors of the elementary field with {\em odd}
number of particles, they coincide with those of Sinh-Gordon at
the ``inverse Yang-Lee point'' (in the notation of ref.\,\cite{FMS}).
In fact, at the self-dual point the minimal form factor $F_{\rm min}$
reduces to the $F_{\rm min}$ of the Sinh-Gordon model at the
``inverse Yang-Lee point'', which was calculated in \cite{FMS}. Then,
using the kinematic residue equations for the $F_{2n-1}$
and comparing it with the ones for Sinh-Gordon \cite{FMS} we obtain that
\EQ
Q^{BD}_{2n-1}\left( B=1,x_1, \dots, x_{2n-1} \right) =
Q^{SG}_{2n-1}\left( B=\frac{2}{3},x_1, \dots, x_{2n-1} \right)
A^{(2n-1)}( x_1, \dots, x_{2n-1} ) \label{eq: BDSG}
\EN
The factor $A^{(2n-1)}$ cancels the poles in the general parametrization
(\ref{para}). Thus, solely from the kinematic residue equation, we obtain that
the form factors of the elementary field of the Bullough-Dodd model at the
self-dual point coincides with the one of the Sinh-Gordon model at the
"inverse Yang-Lee point". In order to finish our proof, we have to
consider eventual additional restrictions coming from the bound state
residue equations. At the self-dual point they become
\EQ
Q_{2n-1}(\omega x,\omega^{-1}x,x_1,\ldots,x_{n-1})\,=\, 0 \,\,\, .
\EN
However this equations is always solved by (\ref{eq: BDSG}), due to the
property of the determinant $A^{(2n-1)}$.

Therefore we have established that in general
\EQ
F_n^{BD}\left( B=1,x_1, \dots, x_{n} \right) \, = \,
F_n^{SG}\left( B=\frac{2}{3},x_1, \dots, x_n \right) \,\,\, ,
\EN
i.e. the BD model reduces at the self-dual point at the Sinh-Gordon model
at the ``inverse Yang-Lee point''.

\resection{Conclusions}

We have derived the recursive equations satisfied by the form factors
of local operators in the BD model. The non-simply laced nature of the
model induces a peculiar analytic structure in the exact $S$-matrix
of the model which, at the self-dual point, coincides with the
$S$-matrix of the Sinh-Gordon model at the ``inverse Yang-Lee point''.
We have proved that the on mass-shell identity between the two theories
extends also off-shell, i.e. the form factors of the elementary field of the
BD model at the self-dual point coincide with those of the elementary field of
the SG model at the ``inverse Yang-Lee point''. A more detailed discussion
of this equivalence, together with general solution of the recursive
equations will be analyzed in \cite{FMS3}.

\section*{Acknowledgments}

Two of us (AF and GM) thank Imperial College for hospitality.
We are grateful to J.L. Cardy, S. Elitzur, D. Olive and A. Schwimmer for
useful discussions.

\newpage

%Begin of Figures
\hs

\vspace{3cm}

\hspace{10cm}

\begin{center}
\begin{picture}(300,160)
\thicklines
\put(80,70){\circle{32}}
\put(80,70){\makebox(0,0){$\Gamma$}}
\put(96,70){\line(1,0){50}}
\put(68.5,58.5){\line(-1,-1){20}}
\put(68.5,81.5){\line(-1,1){20}}
\put(162,70){\circle{32}}
\put(162,70){\makebox(0,0){$\Gamma$}}
\put(173.5,81.5){\line(1,1){20}}
\put(173.5,58.5){\line(1,-1){20}}
\put(200,107){\makebox(0,0){$A$}}
\put(200,33){\makebox(0,0){$A$}}
\put(42,33){\makebox(0,0){$A$}}
\put(42,107){\makebox(0,0){$A$}}
\put(121,60){\makebox(0,0){$A$}}
\end{picture}
\end{center}
\vspace{5mm}
\begin{center}
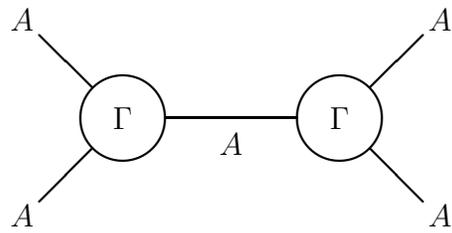
{\bf Figure 1. Bound state pole in the scattering amplitude.}
\end{center}

\end{document}